\documentclass[12pt,a4paper]{article}
\usepackage{amsmath,amsthm,amsfonts,amssymb,bbm}
\usepackage{graphicx,psfrag,subfigure}
\usepackage{cite}
\usepackage{hyperref}

\newcommand{\Id}{\mathbbm{1}}

\newcommand{\Pb}{\mathbbm{P}}

\newcommand{\R}{\mathbb{R}}

\newcommand{\Z}{\mathbb{Z}}
\newcommand{\dx}{\mathrm{d}}

\renewcommand{\Re}{\mathrm{Re}}
\renewcommand{\Im}{\mathrm{Im}}

\DeclareMathOperator*{\Tr}{Tr}

\newtheorem{prop}{Proposition}

\newtheorem{defin}[prop]{Definition}

\newtheorem{cla}[prop]{Claim}

\newtheorem{rem}[prop]{Remark}

\title{From interacting particle systems\\ to random matrices\\ {\normalsize Contribution to StatPhys24 special issue}}

\author{Patrik L. Ferrari\thanks{Institute for Applied Mathematics, University of Bonn, Endenicher Allee 60,\newline 53115 Bonn, Germany; E-mail:~\texttt{ferrari@uni-bonn.de}.\newline This work was supported by the DFG (German Research Foundation) through the SFB~611, project~A12.}}

\date{September 17, 2010}

\begin{document}
\sloppy
\maketitle

\begin{abstract}
In this contribution we consider stochastic growth models in the Kardar-Parisi-Zhang universality class in $1+1$ dimension. We discuss the large time distribution and processes and their dependence on the class on initial condition. This means that the scaling exponents do not uniquely determine the large time surface statistics, but one has to further divide into subclasses.

Some of the fluctuation laws were first discovered in random matrix models. Moreover, the limit process for curved limit shape turned out to show up in a dynamical version of hermitian random matrices, but this analogy does not extend to the case of symmetric matrices. Therefore the connections between growth models and random matrices is only partial.
\end{abstract}

\section{Introduction}
In this paper we discuss results in the Kardar-Parisi-Zhang  (KPZ) universality class of stochastic growth models. We focus on the connections with random matrices occurring in the one-dimensional case. Consider a surface described by a height function $x\mapsto h(x,t)$ with $x\in\R^d$ denoting space and $t\in\R$ being the time variable and subjected to a random dynamics. If the growth mechanism is local and there is a smoothing mechanism providing a deterministic macroscopic growth, then the macroscopic evolution of the interface will be governed by
\begin{equation}
\frac{\partial h}{\partial t}=v(\nabla h)
\end{equation}
where $u\mapsto v(u)$ is the macroscopic growth velocity as a function of the surface slope $u$. In this context, we can also focus on a mesoscopic scale where the random nature of the dynamics is still visible. In the famous paper of Kardar Parisi and Zhang~\cite{KPZ86}, the smoothing mechanism is related with the surface tension and it takes the form $\nu \Delta h$, while the local random dynamics enters as a space-time white noise $\eta$. Moreover, the Taylor expansion of $v$ for small slopes\footnote{The order $0$ and $1$ in the Taylor expansion can be set to be zero by a simple change of (moving) frame.} results into the KPZ equation\footnote{In more than one dimension, $(\nabla h)^2$ should be replaced by $\langle \nabla h, C \nabla h\rangle$ with $C$ a matrix. Then one distinguish the isotropic class, if all the eigenvalues of $C$ have the same sign, and anisotropic class(es) otherwise. For instance, in $d=2$ the surface fluctuation are very different: for anisotropic they are normal distributed in the $\sqrt{\ln(t)}$ scale~\cite{Wol91,PS97} and correlation are the ones of the massless free field~\cite{BF08,BF08b}; for isotropic is it numerical known that growth as $t^\alpha$ for some $\alpha\simeq 0.240$~\cite{TFW92}.}
\begin{equation}\label{eqKPZ}
\frac{\partial h(x,t)}{\partial t}=\nu \Delta h(x,t)+\frac12\lambda (\nabla h(x,t))^2+\eta(x,t),
\end{equation}
where $\lambda=v''(0)\neq 0$ in the non-linear term is responsible for lateral spread of the surface and lack of time reversibility\footnote{When $\lambda=0$ we are in the Edwards-Wilkinson class~\cite{EW82} and the fluctuations are Gaussian with fluctuation exponent $1/4$.}.

From now on we consider the one-dimensional case, $d=1$. Denote by $h_{\rm ma}$ the limit shape,
\begin{equation}
h_{\rm ma}(\xi):=\lim_{t\to\infty}\frac{h(\xi t,t)}{t}.
\end{equation}
The \emph{fluctuation exponent} is $1/3$, the \emph{spatial correlation exponent} is of order $2/3$~\cite{FNS77,BKS85}. This means that the height fluctuations grow in time as $t^{1/3}$ and spatial correlations are $t^{2/3}$, i.e., the rescaled height function at time $t$ around the macroscopic position $\xi$ (at which $h_{\rm ma}$ is smooth)\footnote{However, depending on the initial conditions, (\ref{eqKPZ}) can produce spikes in the macroscopic shape. If one looks at the surface gradient $u=\nabla h$, the spikes of $h$ corresponds to shocks in $u$, and it is known that the shock position fluctuates on a scale $t^{1/2}$. For particular models, properties of the shocks have been analyzed, but mostly for stationary
growth (see~\cite{DJLS93,Fer90,BFS09} and references therein).}
\begin{equation}\label{eqScaling}
h^{\rm resc}_t(u)=\frac{h(\xi t+u t^{2/3},t)-t h_{\rm ma}(\xi+u t^{-1/3})}{t^{1/3}}
\end{equation}
converges to a non-trivial limit process in the $t\to\infty$ limit.
Concerning correlations in space-time, it is known that along special directions the decorrelation occurs only on the macroscopic scale (i.e., with scaling exponent~$1$), while along any other direction the correlations are asymptotically like the spatial correlations at fixed time~\cite{Fer08,CFP10b}. The special directions are the characteristic solutions of the PDE of for the macroscopic height gradient. More precisely, denote by $\xi$ and $\tau$ the macroscopic variables for space and time. Also, let
\begin{equation}
\bar{h}(\xi,\tau):=\lim_{t\to\infty}t^{-1}h(\xi t,\tau t)\quad\textrm{and}\quad u(\xi,\tau):=\partial \bar{h}(\xi,\tau) /\partial\xi.
\end{equation}
Then, $u$ satisfies the PDE
\begin{equation}\label{eqPDE}
\partial u/\partial \tau +a(u)\partial u/\partial\xi=0\quad\textrm{where}\quad
a(u)= -\partial v(u)/\partial u
\end{equation}
with $v$ the macroscopic speed of growth\footnote{In the asymmetric exclusion process explained below, one usually considers the particle density $\rho$ instead of $u$. This is however just a rotation of the frame, since they are simply related by $u=1-2\rho$. The PDE for $\rho$ is the well known Burgers equation.}. The characteristic solutions of (\ref{eqPDE}) are the trajectories satisfying $\partial \xi/\partial \tau=a(u)$ and $\partial u/\partial \tau=0$ (see e.g.~\cite{Ev98,Var04} for more insights on characteristic solutions).

The question is therefore to determine the limit process
\begin{equation}
u\mapsto \lim_{t\to\infty} h^{\rm resc}_t(u) = \textbf{?}
\end{equation}
One might be tempted to think that the scaling exponents are enough to distinguish between classes of models and therefore that the result of our question is independent of the initial condition. However, as we will see, this is not true\footnote{For other observables the relevance of the initial condition was observed already in~\cite{vanB91}.}.

To have an intuition about the relevance of the initial condition, consider the fluctuations of $h(0,t)$ with (a) deterministic initial condition, $h(x,0)=0$ for all $x\in\R$, and (b) random but still macroscopically flat initial condition, $h(x,0)$ a two-sided Brownian motion with $h(0,0)=0$. For this case, the height function $h(0,t)$ is correlated with a neighborhood of $x=0$ of order $t^{2/3}$. In these region (at time $t=0$) for (a) fluctuations are absent while in (b) the fluctuations on the initial condition are of order $t^{1/3}$: this is the same scale as the fluctuations of $h(0,t)$ and therefore the fluctuation laws for (a) and (b) will be different.

How should we proceed to answer to our question? Literally taken is the KPZ equation (\ref{eqKPZ}) ill-defined, because locally one will see a Brownian motion and the problem comes from the square of a white noise (in the non-linear term). However it is possible to give a sense of a solution of the KPZ equation as shown in~\cite{SS10,ACQ10}, and this solution agrees with the one coming from discrete approximations/models (weakly asymmetric simple exclusion process~\cite{BG97}). These works also provide an explicit solution of the finite time one-point distribution for an important initial condition, see~\cite{SS10c} for more explanations.

Another point of view is to see the KPZ equation as one of the models in the KPZ universality class of growth models. By universality it is expected that the limit processes do not depend on the model in the class (but they depend on the type of initial condition). From this perspective, we can take any of the models in the KPZ class and try to obtain the large time limit.

In the rest of the paper we consider one of such model, the totally asymmetric simple exclusion process (TASEP) in which the asymptotic processes have been unraveled. Another model for which analogues results have been determined is the polynuclear growth (PNG) model\footnote{At least half of the limit results described below were first obtained for the PNG model.}. In particular, we will discuss which limit distributions/processes also appears in random matrix theory and when the connection is only partial.

\section{TASEP}
The totally asymmetric simple exclusion process (TASEP) in continuous time is the simplest non-reversible interacting stochastic particle system.
In TASEP particles are on the lattice of integers, $\Z$, with at most one particle at each site (exclusion principle).
\begin{figure}
\begin{center}
\psfrag{rate 1}[l]{rate 1}
\includegraphics[height=1.5cm]{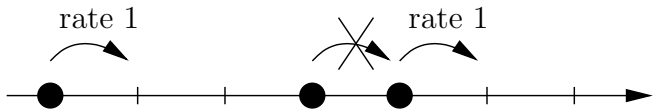}
\caption{Dynamics of the TASEP. Particles jump with rate one on the right site, but under the constraint that the site is empty.}
\label{FigTASEPdyn}
\end{center}
\end{figure}
The dynamics is defined as follows. Particles jump to the neighboring right site with rate $1$ provided that the site is empty. Jumps are independent of each other and take place after an exponential waiting time with mean $1$, which is counted from the time instant when the right neighbor site is empty, see Figure~\ref{FigTASEPdyn}.

\begin{figure}
\begin{center}
\psfrag{rate 1}[cb]{rate 1}
\includegraphics[height=4cm]{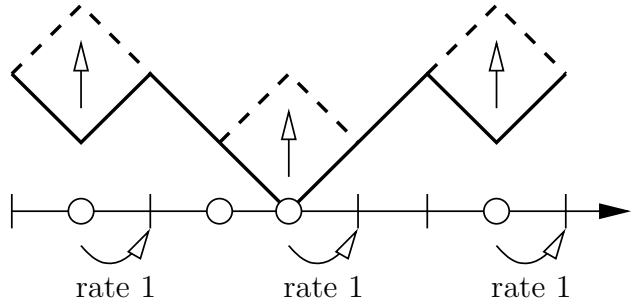}
\caption{Height configuration (solid line) associated with a particle (circles) configuration. There are three particles which can jump and the corresponding evolution of the height function is illustrated by the dashed profiles.}
\label{FigTASEPheight}
\end{center}
\end{figure}

Denote by $\eta_x(t)$ the occupation variable of site $x\in\Z$ at time $t\geq 0$, i.e., $\eta_x(t)$ is $1$ if there is a particle and $0$ if the site is empty. TASEP configurations are in bijection with the surface profile defined by setting the origin \mbox{$h(0,0)=0$} and the discrete height gradient to be $1-2\eta_x(t)$. If we denote by $N_t$ the number of particles which have crossed the bond $0$ to $1$ during the time span $[0,t]$, then the height function is given by
\begin{equation}
h(x,t)=\left\{
\begin{array}{ll}
  2N_t+\sum_{y=1}^x(1-2\eta_y(t)), & \textrm{for }x\geq 1, \\
  2N_t, & \textrm{for }x=0, \\
  2N_t-\sum_{y=x+1}^0(1-2\eta_y(t)), & \textrm{for }x\leq -1.
\end{array}
\right.
\end{equation}
as illustrated in Figure~\ref{FigTASEPheight}.

Let us verify that TASEP belongs to the KPZ universality class. Under hydrodynamical scaling, the particle density $\rho$ evolves according to the Burgers equation $\partial_t\rho +\partial_x(\rho(1-\rho))=0$. Thus, we have a deterministic limit shape. The second requirement, the locality of the growth dynamics is obviously satisfied. Finally, the speed of growth $v$ of the interface is twice the current density, which is given by $\rho(1-\rho)$. Being the gradient $u=1-2\rho$ it follows that $v(u)=(1-u^2)/2$, which implies $v''(u)=-1\neq 0$.

Now we discuss some of the large time results for the TASEP height function\footnote{Most of the results has been first computed for particle positions and the statements described below are obtained by a geometric transformation.}. We consider two non-random initial conditions generating a curved and a flat macroscopic shape. The limit processes will be called the Airy$_2$ and Airy$_1$ processes, which are defined in Appendix~\ref{AppAirys}.

\subsection{TASEP with step initial condition}
\begin{figure}
\begin{center}
\includegraphics[height=3cm]{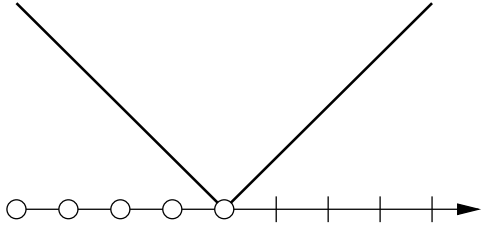}
\caption{Height configuration for step initial conditions.}
\label{FigTASEPICdroplet}
\end{center}
\end{figure}

Consider the initial condition $\eta_x(0)=1$ for $x\leq 0$ and $\eta_x(0)=0$ for $x\geq 1$, see Figure~\ref{FigTASEPICdroplet}. This is called \emph{step-initial condition}.

The macroscopic limit shape for this initial condition is a parabola continued by two straight lines:
\begin{equation}
h_{\rm ma}(\xi)=
\left\{
\begin{array}{ll}
  \frac12(1+\xi^2), & \textrm{for }|\xi|\leq 1,\\[0.5em]
  |\xi|, & \textrm{for }|\xi|\geq 1.
\end{array}
\right.
\end{equation}
From this we have the scaling\footnote{With respect to (\ref{eqScaling}) we adjusted the coefficients to avoid having them in the asymptotic process.}
\begin{equation}
h^{\rm resc}_t(u):=\frac{h(2u(t/2)^{2/3},t)-\left(t/2+u^2(t/2)^{1/3}\right)}{-(t/2)^{1/3}}.
\end{equation}

The large time results for the rescaled height function $h^{\rm resc}_t$ are the following. First, for the one-point distribution~\cite{Jo00b}
\begin{equation}
\lim_{t\to\infty}\Pb\left(h^{\rm resc}_t(0)\leq s\right)=F_2(s),
\end{equation}
where $F_2$ is known as the GUE Tracy-Widom distribution, first discovered in random matrices~\cite{TW94} (see Section~\ref{sectRM}). Moreover, concerning the joint distributions, it is proven~\cite{BFS07,BF07,Jo03b} that (in the sense of finite-dimensional distribution\footnote{In~\cite{Jo03b} Johansson the process was studied in a slightly different cut, but because of slow-decorrelation~\cite{Fer08,CFP10b} the present result can be proven from it. Remark also that the convergence in~\cite{Jo03b} is in a stronger sense.})
\begin{equation}
\lim_{t\to\infty} h^{\rm resc}_t(u)={\cal A}_2(u),
\end{equation}
where ${\cal A}_2$ is called the Airy$_2$ process, first discovered in the PNG model by Pr\"ahofer and Spohn~\cite{PS02} (see Appendix~\ref{AppAirys})\footnote{The height function of TASEP in discrete time with parallel update and step initial condition is the same as the arctic line in the Aztec diamond for which the Airy$_2$ process was obtained by Johansson in~\cite{Jo03}. Extensions to process on space-like path for the PNG model was made in~\cite{BO04}. Tagged particle problem was studied in~\cite{SI07}, extension to space-like paths in~\cite{BF07,BFS07b} and to any space-time paths except characteristic line in~\cite{Fer08,CFP10b}.}. In particular, the Airy$_2$ process is stationary, locally looks like a Brownian motion and has correlations decaying slow: like $u^{-2}$ (see Figure~\ref{FigCovariances}).

By universality it is expected that the Airy$_2$ process describes the large time surface statistics for initial conditions\footnote{Also for random initial conditions as in the case of Bernoulli-$\rho_-$ on $\Z_-$ and Bernoulli-$\rho_+$ on $\Z_+$, $\rho_->\rho_+$, see~\cite{BC09,PS01}.} generating a smooth curved macroscopic shape for models in the KPZ class. This happens when the characteristic lines for space-time points on the curved limit shape go all together at a single point at time $t=0$ (for the TASEP and PNG are straight lines back to the origin).

\subsection{TASEP with flat initial condition}
\begin{figure}
\begin{center}
\includegraphics[height=1cm]{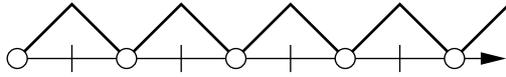}
\caption{Height configuration for a deterministic and flat initial conditions.}
\label{FigTASEPICflat}
\end{center}
\end{figure}

The second type of non-random initial condition we discuss here is called \emph{flat-initial condition}. In terms of TASEP particles, it is given by $\eta_x(0)=1$ for $x$ even and $\eta_x(0)=0$ for $x$ odd, see Figure~\ref{FigTASEPICflat}.

\begin{figure}
\begin{center}
\psfrag{u}[c]{$u$}
\psfrag{g1}[r]{$g_1(u)$}
\psfrag{g2}[r]{$g_2(u)$}
\includegraphics[height=6cm]{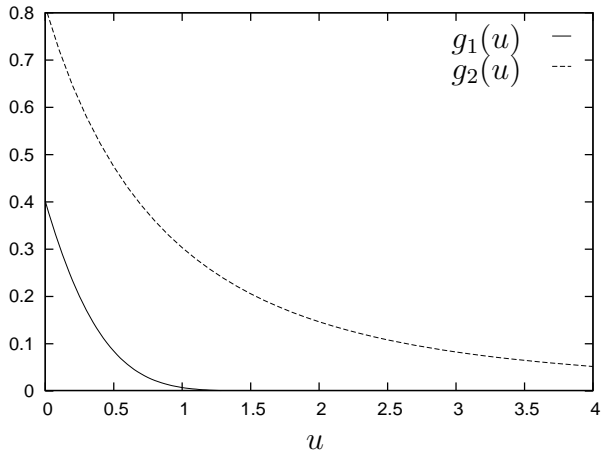}
\caption{Covariance $g_2(u)={\rm Cov}({\cal A}_2(u),{\cal A}_2(0))$ of the Airy$_2$ process (dashed line) and $g_1(u)={\rm Cov}({\cal A}_1(u),{\cal A}_1(0))$ of the Airy$_1$ process (solid line). One clearly sees the difference of behavior: $g_2(u)\simeq 2 u^{-2}$ for $u\gg 1$, while $g_1(u)$ goes to zero super-exponentially fast.}
\label{FigCovariances}
\end{center}
\end{figure}

The macroscopic limit shape is very simple, $h_{\rm ma}(\xi)=\frac12$, so that the rescaled height function becomes
\begin{equation}
h^{\rm resc}_t(u):=\frac{h(2u t^{2/3},t)-t/2}{-t^{1/3}}.
\end{equation}
In the large time limit, the one-point distribution of $h^{\rm resc}_t$ is given by\footnote{For the geometric case corresponding to discrete time TASEP this result was proven by Baik and Rains in~\cite{BR99}.}
\begin{equation}
\lim_{t\to\infty}\Pb\left(h^{\rm resc}_t(0)\leq s\right)=F_1(2s),
\end{equation}
where $F_1$ is known as the GOE Tracy-Widom distribution, first discovered in random matrices~\cite{TW96} (see Section~\ref{sectRM}). Moreover, as a process, it was discovered by Sasamoto~\cite{Sas05,BFPS06} and it is proven that (in the sense of finite-dimensional distribution)
\begin{equation}
\lim_{t\to\infty} h^{\rm resc}_t(u)={\cal A}_1(u),
\end{equation}
where ${\cal A}_1$ is called the Airy$_1$ process (see Appendix~\ref{AppAirys}). In particular, the Airy$_1$ process is stationary, it behaves locally like a Brownian motion, but unlikely for the Airy$_2$ process, the decorrelations decay superexponentially fast (see Figure~\ref{FigCovariances}), see the review~\cite{Fer07} for more information and references.

The Airy$_1$ process is expected to describe the large time surface behavior for non-random initial conditions generating a straight limit shape for models in the KPZ class. Unlike for the curved limit shape, the characteristic lines\footnote{For density $1/2$ the characteristic lines are all lines parallel to the time axis. For density $\rho$, they have the form $x=x_0+(1-2\rho)t$.} for space-time points for flat limit shape do not join at initial time. This fact is at the origin of (a) the different fluctuation behavior between curved and flat and (b) the difference between random flat and non-random flat.

\subsection{TASEP with stationary initial condition}
The only translation invariant stationary measure for continuous time TASEP are Bernoulli product measures with parameter $\rho$, $\rho\in [0,1]$, which is the density of particles~\cite{Lig76}. The cases $\rho\in\{0,1\}$ are degenerate and nothing happens, so consider a fixed $\rho\in (0,1)$. The height function at time $0$ is a two-sided random walk with $h(0,0)=0$ and $\Pb(h(x+1,0)-h(x,0)=1)=1-\rho$ and $\Pb(h(x+1,0)-h(x,0)=-1)=\rho$.

Unlike the deterministic initial conditions we need to consider the regions where the height function is non-trivially correlated with $h(0,0)$. The reason is that for the regions at time $t$ which are correlated with $h(\alpha t,0)$, $\alpha\neq 0$, the dynamical fluctuations (of order $t^{1/3}$) and are dominated by the fluctuations in the initial condition (of order $t^{1/2}$). The correlations in TASEP are carried by second-class particles, which move with speed $1-2\rho$. As predicted by the KPZ scaling, the limit
\begin{equation}
\lim_{t\to\infty}\frac{h((1-2\rho)t+u t^{2/3},t)-(1-2\rho(1-\rho))t}{t^{1/3}}
\end{equation}
exists. The one-point distribution was derived in the PNG model with external source~\cite{BR00} and for TASEP in~\cite{FS05a} (correctly conjectured using universality in~\cite{PS01}). The extension to joint distributions is worked out in~\cite{BFP09}. So-far no connections with random matrices for this initial condition is known. Therefore we do not enter in further details.

\section{Random Matrices}\label{sectRM}
We said that $F_1$ and $F_2$ appeared first in random matrices. We will explain it below and discuss whether the Airy processes also show up for random matrix models.

\subsection{Hermitian matrices}
The Gaussian Unitary Ensemble (GUE) of random matrices consists of Hermitian matrices $H$ of size $N\times N$ distributed according to the probability measure\footnote{The scaling of this paper is such that the asymptotic density of eigenvalues remains bounded, the macroscopic variable is $N$ and scaling exponents are easily compared with KPZ. In the literature there are other two standard normalization constants, which are just a rescaling of eigenvalues. The first one consists of replacing $1/(2N)$ by $N$: this is appropriate if one study spectral properties, since in the large $N$ limit the spectrum remains bounded. The second is to replace $1/(2N)$ by $1$ so that the measure does not depend on $N$: this is most appropriate if one looks at eigenvalues' minors.}
\begin{equation}\label{eqDefGUE}
p^{\rm GUE}(H)\dx H=\frac{1}{Z_N}\exp\left(-\frac{1}{2N}\Tr(H^2)\right)\dx H,
\end{equation}
where $\dx H=\prod_{i=1}^N \dx H_{i,i}\prod_{1\leq i<j\leq N}\dx \Re(H_{i,j})\dx \Im(H_{i,j})$ is the reference measure (and $Z_N$ the normalization constant).

\begin{figure}
\begin{center}
\psfrag{F}[]{}
\psfrag{ 0.5}[r]{$0.5$}
\psfrag{ 0.4}[r]{$0.4$}
\psfrag{ 0.3}[r]{$0.3$}
\psfrag{ 0.2}[r]{$0.2$}
\psfrag{ 0.1}[r]{$0.1$}
\psfrag{ 0}[r]{$0$}
\psfrag{ 00}[c]{$0$}
\psfrag{-4}[c]{$-4$}
\psfrag{-2}[c]{$-2$}
\psfrag{ 2}[c]{$2$}
\psfrag{ 4}[c]{$4$}
\psfrag{s}[b]{$s$}
\psfrag{b1}[r]{$F_1'(s)$}
\psfrag{b2}[r]{$F_2'(s)$}
\includegraphics[height=6cm]{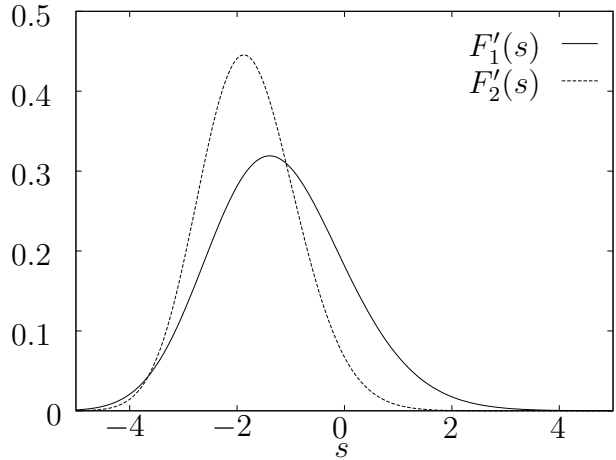}
\caption{Dashed line: the densities of the GUE Tracy-Widom distribution; solid line: the GOE Tracy-Widom distribution density.}
\label{FigF12}
\end{center}
\end{figure}

Denote by $\lambda_{N,\rm max}^{\rm GUE}$ the largest eigenvalue of a $N\times N$ GUE matrix. Then Tracy and Widom in~\cite{TW94} proved that the asymptotic distribution of the (properly rescaled) largest eigenvalue is $F_2$ (see Figure~\ref{FigF12}):
\begin{equation}
\lim_{N\to\infty} \Pb\left(\frac{\lambda^{\rm GUE}_{N,\rm max}-2N}{N^{1/3}}\leq s\right)=F_2(s).
\end{equation}

The parallel between GUE and TASEP with step initial condition goes even further. In 1962 Dyson~\cite{Dys62} introduced a matrix-valued Ornstein-Uhlenbeck process which is now called Dyson's Brownian Motion (DBM). For hermitian matrices, GUE DBM is the stationary process on matrices $H(t)$ whose evolution is governed by
\begin{equation}\label{eqDefDBM2}
\dx H(t)=-\frac{1}{2N}H(t)\dx t+\dx B(t)
\end{equation}
where $\dx B(t)$ is a (hermitian) matrix-valued Brownian motion. More precisely, the entries $B_{i,i}(t)$, $1\leq i\leq N$, $\Re(B_{i,j})(t)$ and $\Im(B_{i,j})(t)$, $1\leq i<j\leq N$, perform independent Brownian motions with variance $t$ for diagonal terms and $t/2$ for the remaining entries. Denote by $\lambda^{\rm GUE}_{N,\rm max}(t)$ the largest eigenvalue at time $t$ (when started from the stationary measure (\ref{eqDefGUE})). Its evolution is, in the large $N$ limit, governed by the Airy$_2$ process:
\begin{equation}\label{eqDBM2}
\lim_{N\to\infty}\frac{\lambda^{\rm GUE}_{N,\rm max}(2u N^{2/3})-2N}{N^{1/3}}={\cal A}_2(u).
\end{equation}

Thus we have seen that the connection between GUE and TASEP extends to the process\footnote{This connection extends \emph{partially} to the evolution of minors, see~\cite{FF10,ANvM10}.}.

\subsection{Symmetric matrices}
The Gaussian Orthogonal Ensemble (GOE) of random matrices consists of symmetric matrices $H$ of size $N\times N$ distributed according to
\begin{equation}\label{eqDefGOE}
p^{\rm GOE}(H)\dx H=\frac{1}{Z_N}\exp\left(-\frac{1}{4N}\Tr(H^2)\right)\dx H,
\end{equation}
where $\dx H=\prod_{1\leq i\leq j\leq N}\dx H_{i,j}$ is the reference measure (and $Z_N$ the normalization constant).

Denote by $\lambda^{\rm GOE}_{N,\rm max}$ the largest eigenvalue of a $N\times N$ GOE matrix. The asymptotic distribution of the (properly rescaled) largest eigenvalue is $F_1$ (see Figure~\ref{FigF12})~\cite{TW96}:
\begin{equation}
\lim_{N\to\infty} \Pb\left(\frac{\lambda^{\rm GOE}_{N,\rm max}-2N}{N^{1/3}}\leq s\right)=F_1(s).
\end{equation}

DBM is defined also for symmetric matrices: GOE DBM is the stationary process on matrices $H(t)$ whose evolution is governed by
\begin{equation}\label{eqDefDBM1}
\dx H(t)=-\frac{1}{4N}H(t)\dx t+\dx B(t)
\end{equation}
where $\dx B(t)$ is a symmetric matrix-valued Brownian motion. More precisely, the entries $B_{i,j}(t)$, $1\leq i\leq j\leq N$, perform independent Brownian motions with variance $t$ for diagonal terms and $t/2$ for the remaining entries.

\begin{figure}
\begin{center}
\includegraphics[height=7cm]{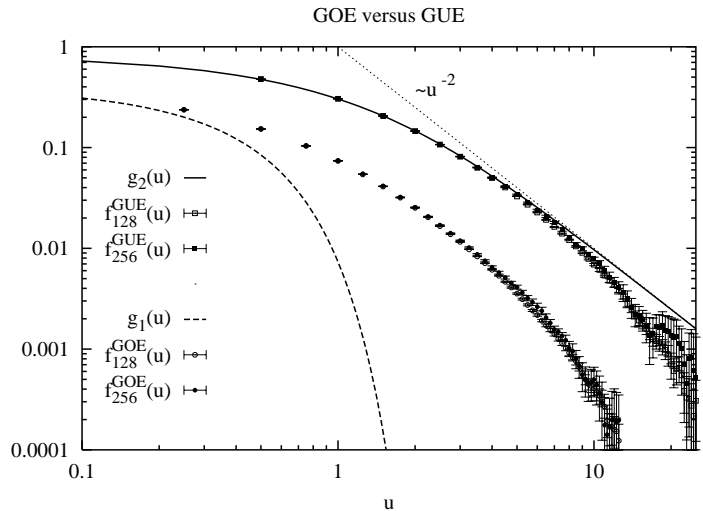}
\caption{Log-log plot of the rescaled correlation functions for GOE and GUE. For $k=1,2$, we denote $g_k={\rm Cov}({\cal A}_k(u),{\cal A}_k(0))$ and $f_{N}^{\rm GUE}(u)$ (resp.\ $f_{N}^{\rm GOE}(u)$) is the covariance of the GUE (resp.\ GOE) largest eigenvalues of a $N\times N$ matrix rescaled as in (\ref{eqDBM2}) (resp.\ (\ref{eqDBM1})).}
\label{FigNotAiry1}
\end{center}
\end{figure}

We consider now the evolution of the largest eigenvalue $\lambda^{\rm GOE}_{N,\rm max}(t)$ (when started from the stationary distribution (\ref{eqDefGOE})). By analogy with the GUE case, one might guess that in the large $N$ limit the limit process of a properly rescaled $\lambda^{\rm GOE}_{N,\rm max}(t)$ is the Airy$_1$ process. However, as shown numerically in~\cite{BFP08}, this is not the case. To see this, we considered the scaling (\ref{eqDBM1}) where the coefficients are chosen such that the variance at $u=0$ is the same as the variance of the Airy$_1$ process, and the covariance for $|u|\ll 1$ coincide at first order with the covariance of ${\cal A}_1$. The large time limit of the rescaled largest eigenvalue is denoted by ${\cal B}_1$. Comparing the covariances as shown in Figure~\ref{FigNotAiry1} we conclude that ${\cal A}_1\neq{\cal B}_1$:
\begin{equation}\label{eqDBM1}
\lim_{N\to\infty}\frac{\lambda^{\rm GOE}_{N,\rm max}(8u N^{2/3})-2N}{2N^{1/3}}=:{\cal B}_1(u) \neq {\cal A}_1(u).
\end{equation}
Thus the connection between GOE and TASEP \emph{does not extend} to multi-time distributions\footnote{The connection extends to the top eigenvalues too~\cite{Fer04}, but not still at a fixed time only.}.

\section{Conclusion}
We saw that large time fluctuations in KPZ growth models depends on the initial conditions. By analyzing special models we determine the limiting distributions and processes, which by universality should be the same for whole KPZ universality class. The conjecture~\cite{PS00} is that when the limit shape is curved one gets the Airy$_2$ process, when the limit shape is flat one has to further distinguish according to roughness exponent $\alpha$ of the initial condition ($|h(x,0)-h(0,0)|\sim |x|^\alpha$). For $\alpha=0$ one expects the Airy$_1$ process, for $\alpha=1/2$ the result from stationary initial condition. Finally, for $0<\alpha<1/2$, at the characteristics it should still be Airy$_1$ but there should be a region (away of order $t^{1/(3\alpha)}$ from the characteristic lines) with a different (yet unknown) process, which depends on the exponent $\alpha$.

\begin{figure}
\subfigure[]{
\psfrag{KPZ}[c]{\small KPZ class}
\psfrag{CurvedKPZ}[c]{\small Curved shape}
\psfrag{TASEP}[c]{\small TASEP, step IC}
\psfrag{RM}[c]{\small Random matrices}
\psfrag{HermitianRM}[c]{\small Hermitian}
\psfrag{GUE}[c]{\small GUE matrices}
\psfrag{F2}[c]{\small $F_2$ distribution}
\psfrag{Airy2}[c]{\small ${\cal A}_2$ process}
\psfrag{T}[l]{\small $t\to\infty$}
\psfrag{N}[r]{\small $N\to\infty$}
\includegraphics[width=0.5\textwidth]{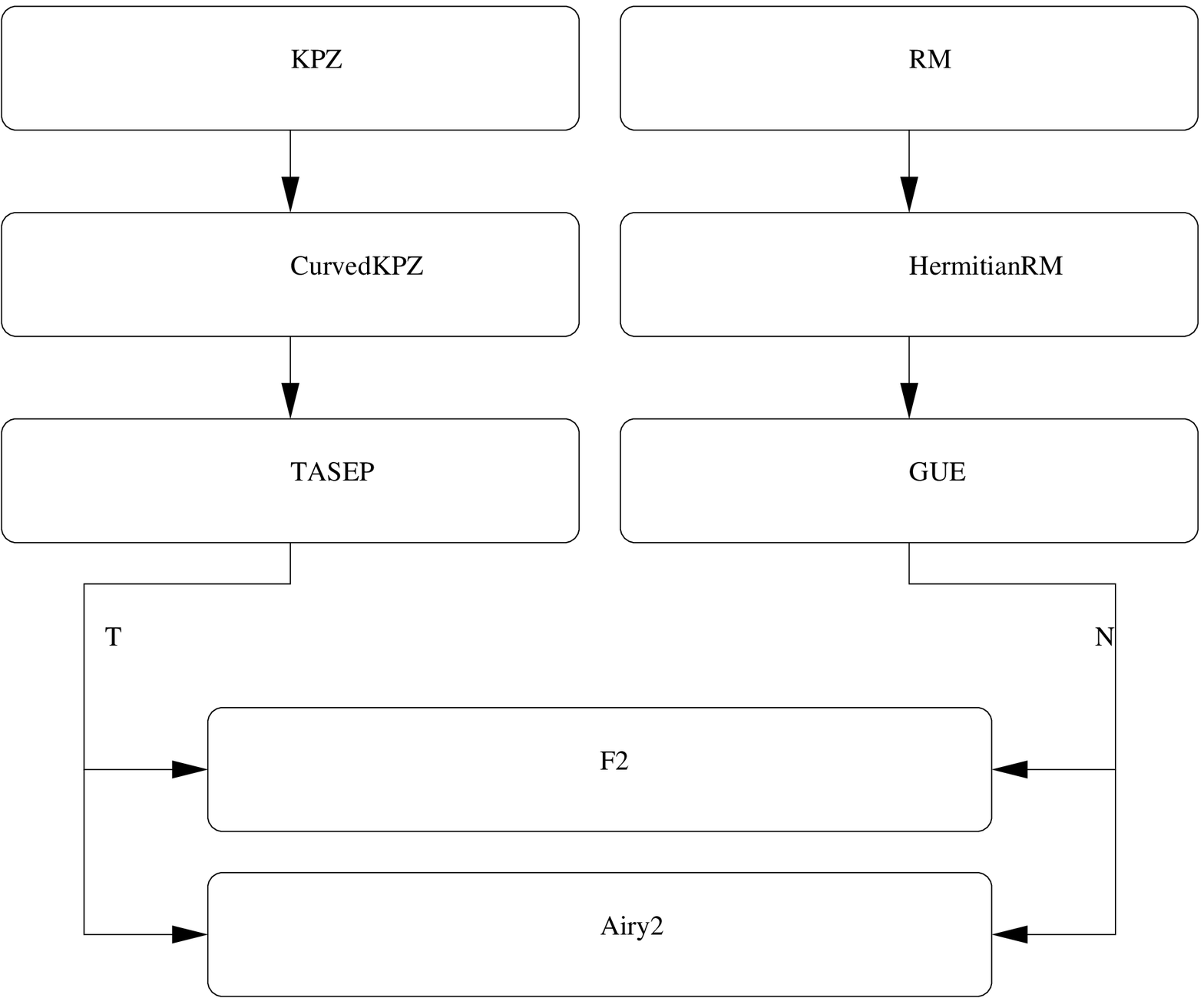}
}
\hfill
\subfigure[]{
\psfrag{KPZ}[c]{\small KPZ class}
\psfrag{FlatKPZ}[c]{\small Flat shape}
\psfrag{TASEP}[c]{\small TASEP, flat IC}
\psfrag{RM}[c]{\small Random matrices}
\psfrag{SymmetricRM}[c]{\small Symmetric}
\psfrag{GOE}[c]{\small GOE matrices}
\psfrag{F1}[c]{\small $F_1$ distribution}
\psfrag{Airy1}[c]{\small ${\cal A}_1$ process}
\psfrag{unknown}[c]{\small Unknown}
\psfrag{neq}[c]{\tiny $\neq$}
\psfrag{T}[l]{\small $t\to\infty$}
\psfrag{N}[r]{\small $N\to\infty$}
\includegraphics[width=0.5\textwidth]{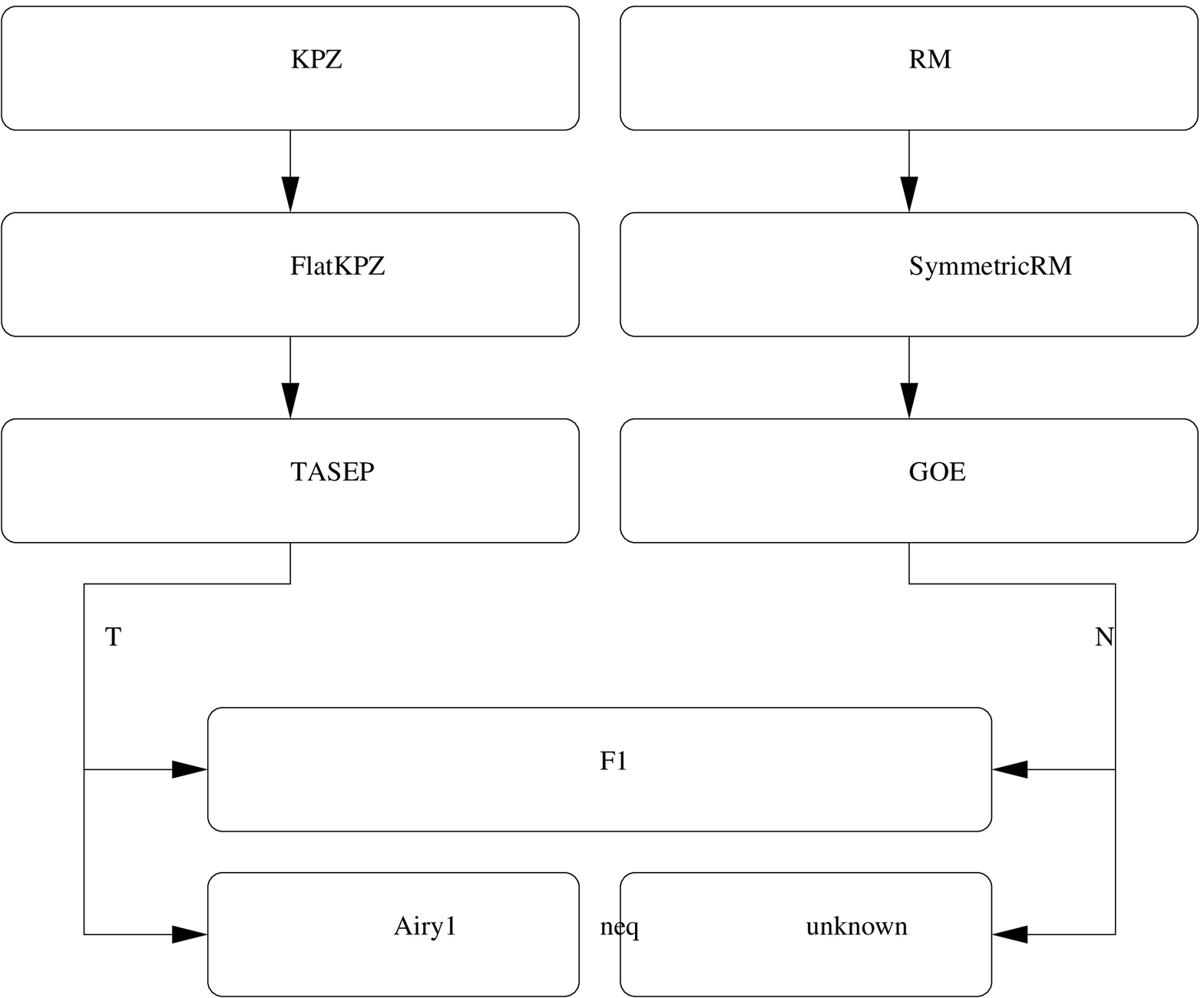}
}
\caption{Comparison: non-random initial conditions for KPZ vs Random matrices: (a) curved limit shape vs Hermitian matrices, (b) flat limit shape vs symmetric matrices.}
\label{FigScheme}
\end{figure}

\noindent The two scheme in Figure~\ref{FigScheme} resume the main message of this contribution:
\begin{itemize}
 \item[(a)] TASEP with step initial condition is a representant for models in the KPZ class with curved limit shape. For this model the large time limit distribution is $F_2$ and the limit process the Airy$_2$ process, ${\cal A}_2$. The GUE (with DBM dynamics) is a representant of random matrices with hermitian symmetries and also for this model the $F_2$ distribution and the process ${\cal A}_2$ arises in the limit of large matrices ($N\to\infty$).
 \item[(b)] TASEP with flat initial condition is one of the KPZ growth models with deterministic initial conditions and straight limit shape. The long time fluctuations are governed by the $F_1$ distributions and the limit process is the Airy$_1$ process. On the random matrix side, GOE (with DBM dynamics) is a representant for symmetric random matrices and $F_1$ is the distribution of its rescaled largest eigenvalue as $N\to\infty$. However, it is not the Airy$_1$ process which describes its dynamical extension.
\end{itemize}

The interested reader is referred to~\cite{FS10,KK08} for more insights about random matrices and stochastic growth models. Therein a guide of literature is provided. Also, two summer schools lecture notes on the subject are available~\cite{Spo05,Jo05}.

On the experimental side, recently Takeuchi built up a very nice experimental set-up in which the theoretical prediction (scaling exponents, one-point distributions and covariance) have been verified with very good agreement both for curved limit shape~\cite{TS10} and flat limit shape~\cite{Tak10}.

\appendix
\section{Definition of the limit processes}\label{AppAirys}
In this short appendix we give the definitions of the Airy$_1$ and Airy$_2$ processes. More information about these processes like properties and references can be found for example in the reviews~\cite{FS10,Fer07}.
\begin{defin}
The {Airy$_1$ process} ${\mathcal{A}}_{\rm 1}$ is the process with $m$-point joint distributions at
\mbox{$u_1< u_2< \ldots < u_m$} given by the Fredholm determinant
\begin{equation}\label{eqF1}
\Pb\Big(\bigcap_{k=1}^m\{{\mathcal{A}}_{\rm 1}(u_k)\leq s_k\}\Big)=
\det(\Id-\chi_s K_1\chi_s)_{L^2(\{u_1,\ldots,u_m\}\times\R)}
\end{equation}
where $\chi_s(u_k,x)=\Id(x>s_k)$ and the kernel $K_1$ is given by
\begin{equation}\label{eqKernelF1}
\begin{aligned}
K_1(u,s;u',s')&=-\frac{1}{\sqrt{4\pi (u'-u)}}\exp\left(-\frac{(s'-s)^2}{4 (u'-u)}\right) \mathbf{1}(u<u') \\
& +\mathrm{Ai}(s+s'+(u'-u)^2) \exp\left((u'-u)(s+s')+\frac23(u'-u)^3\right)
\end{aligned}
\end{equation}
\end{defin}

\begin{defin}
The {Airy$_2$ process} ${\mathcal{A}}_{\rm 2}$ is the process with $m$-point joint distributions at
\mbox{$u_1< u_2< \ldots < u_m$} given by the Fredholm determinant
\begin{equation}\label{eqF2}
\Pb\Big(\bigcap_{k=1}^m\{{\mathcal{A}}_{\rm 2}(u_k)\leq s_k\}\Big)=
\det(\Id-\chi_s K_2\chi_s)_{L^2(\{u_1,\ldots,u_m\}\times\R)}
\end{equation}
where $\chi_s(u_k,x)=\Id(x>s_k)$ and $K_2$ is the extended Airy kernel given by
\begin{equation}\label{eqKernelF2}
K_2(u,s;u',s')=\left\{\begin{array}{ll}
\int_{\mathbb{R}_+}\mathrm{d}\lambda e^{(u'-u)\lambda}\mathrm{Ai}(x+\lambda)\mathrm{Ai}(y+\lambda),&u\geq u',\\[0.5em]
-\int_{\mathbb{R}_-}\mathrm{d}\lambda e^{(u'-u)\lambda}\mathrm{Ai}(x+\lambda)\mathrm{Ai}(y+\lambda),&u<u'.
\end{array}\right.
\end{equation}
\end{defin}


\end{document}